\documentclass[aps,prc,a4paper,nofootinbib,preprintnumbers,superscriptaddress,showpacs,showkeys]{revtex4}

\usepackage{amsmath}
\usepackage{amssymb}
\usepackage{bm}
\usepackage{graphicx}
\usepackage{color}

\newcommand{\be}{\begin{equation}}
\newcommand{\ee}{\end{equation}}
\newcommand{\ba}{\begin{eqnarray}}
\newcommand{\ea}{\end{eqnarray}}

\begin{document}

\title{Classical model for diffusion and thermalization of heavy quarks in a hot medium:\\
memory and out-of-equilibrium effects}

\author{Marco Ruggieri}\email{ruggieri@lzu.edu.cn}
\affiliation{School of Nuclear Science and Technology, Lanzhou University, 222 South Tianshui Road, Lanzhou 730000, China}
\author{Marco Frasca}
\affiliation{Via Erasmo Gattamelata, 00176 Rome, Italy}
\author{Santosh Kumar Das}
\affiliation{School of Physical Science, Indian Institute of Technology Goa, Ponda-403401, Goa, India}


\begin{abstract}
We consider a simple model for the diffusion of heavy quarks in a hot bath, modeling the latter by an ensemble
of oscillators distributed accorded to either a thermal distribution or to an
out-of-equilibrium distribution with a saturation scale. Within this model it is easy to introduce
memory effects  by changing the distribution of oscillators:
we model these by introducing a gaussian distribution, $dN/d\omega$,
which can be deformed continuously from a $\delta-$function giving a Markov dissipation to a broad kernel with memory.
Deriving the equation of motion of the heavy quark in the bath we remark how 
dissipation comes out naturally as an effect of the back-reaction 
on the bath of oscillators.  Moreover, the exact solution of this equation allows for the definition of a thermalization time
as the time necessary to remove any memory of the initial condition.
We find that the broadening the dissipative kernel while keeping the coupling
fixed lowers the thermalization time. We also derive
the fluctuation-dissipation theorem for the bath, and use this to estimate the kinematic regime in which
momentum diffusion of the heavy quark dominates over drift: we find that diffusion is more important
as long as $K_0/{\cal E}$ is small, where $K_0$ and ${\cal E}$ denote the initial energy of the heavy quark
and the average energy of the bath respectively.  
\end{abstract}



\maketitle

\section{Introduction}

 The description of high energy collisions as the interaction of sheets of color-glass-condensate (CGC)
\cite{McLerran:1993ni,McLerran:1993ka,McLerran:1994vd,Gelis:2010nm,Iancu:2003xm,McLerran:2008es,
Iancu:2000hn,Gelis:2012ri},
is one of the most intriguing approach to the study of relativistic proton-proton (pp),
proton-nucleus (pA) and nucleus-nucleus (AA) collisions
at the Relativistic Heavy Ion Collider (RHIC) and Large Hadron Collider (LHC) energies.
One of the predictions of the CGC effective theory is the formation of
strong gluon fields in the forward light cone
named the Glasma
\cite{Kovner:1995ja,Kovner:1995ts,Gyulassy:1997vt,Lappi:2006fp,Fukushima:2006ax,
Fries:2006pv,Chen:2015wia,Fujii:2008km,Krasnitz:2000gz,Krasnitz:2003jw,Krasnitz:2001qu,Ruggieri:2017ioa} that
consists of out-of-equilibrium longitudinal color-electric and color-magnetic fields;
these are characterized by large gluon occupation number,
$A_\mu^a \simeq 1/g$ with $g$ representing the strong coupling, 
so they can be described by classical field theory namely the Classical Yang-Mills  (CYM) theory.
Heavy quarks are excellent probes of the system created in high energy nuclear collisions,
both for the pre-equilibrium part and for the thermalized quark-gluon plasma (QGP), see
\cite{Rapp:2018qla, Aarts:2016hap, Greco:2017rro,Das:2015ana, Das:2013kea,Das:2016cwd,
Das:2017dsh,Das:2015aga,Beraudo:2015wsd,Xu:2015iha,Ozvenchuk:2017ojj,
Prino:2016cni,Andronic:2015wma,Mrowczynski:2017kso,Ruggieri:2018rzi,Ruggieri:2018ies,Sun:2019fud,
Cao:2018ews,Dong:2019unq,Svetitsky:1987gq,Moore:2004tg,Scardina:2017ipo,vanHees:2007me,vanHees:2005wb,
 Alberico:2013bza,Cao:2013ita,Schmidt:2014zpa,Wesp:2014xpa,He:2013zua} 
and references therein.
Their formation time is very small in comparison with the one of light quarks,
 $\tau_\mathrm{form} \leq 0.1$ fm/c for both charm and beauty quark. 
Because heavy quarks are produced
immediately after the collision, they can propagate in the evolving Glasma fields and probe its evolution.
Propagation of heavy quarks in the Glasma has been studied previously within 
the Fokker-Planck equation  \cite{Mrowczynski:2017kso} as well as the Wong equations coupled to the
evolving Glasma background field \cite{Ruggieri:2018rzi,Ruggieri:2018ies}.
The common result of these references has been to find a diffusion of heavy quarks in the dense
gluon medium, very similar to that in a thermal medium; on the other hand,
a substantial drift has not been observed in \cite{Ruggieri:2018rzi,Ruggieri:2018ies}
in the kinematic regime examined there, which paved the way for a possible solution
of the $R_{AA}/v_2$ puzzle of heavy quarks in Pb-Pb collisions \cite{Sun:2019fud}. 

The aim of the present work is to use a simple
classical mechanics model to describe the motion of the heavy quark momentum in a dense and hot bath,
that has been used extensively in connection to dissipation in the quantum Brownian motion,
see for example \cite{RUSSO,QHO,QHO2,CHAOS} as well as references therein. 
The advantage of this model, despite its simplicity, is that it has the potential to describe several aspects that 
might be important to understand the drift and diffusion of these quarks, among them the propagation in a bath out of equilibrium
like the Glasma, as well as the memory effects in the dissipation kernel.
In the particular context of high energy nuclear collisions,  memory 
has started to attract some interest \cite{Greiner:1994xm, Murase:2013tma, Kapusta:2014dja,Schmidt:2014zpa, Wesp:2014xpa}
as it might affect the 
 dilepton yields~\cite{Schenke:2006uh} as well as radiative energy loss~\cite{Michler:2009dy}. 
In this model, the bath is described as a system of harmonic oscillators;
initial conditions for the oscillators are randomly assigned, with expectation values for the squared
momenta and positions in agreement with the equipartition theorem in case of a bath in thermal equilibrium,
or to the energy per particle in case the bath is out of equilibrium.
To model an out-of-equilibrium bath we will use a distribution inspired by CGC models
which is flat below a certain energy scale, the saturation scale $Q_s$, then rapidly drops to zero
\cite{Scardina:2014gxa,Blaizot:2013lga,Blaizot:2011xf}. 
The coupling of the bath to the heavy quark is pretty simple, $H_\mathrm{int}\propto Q \sum_k x_k$,
where $Q$ is the generalized coordinate of the heavy quark and $x_k$ denotes the position
of the $k^\mathrm{th}$ oscillator; this simple coupling is enough to emphasize that 
in order to achieve energy loss during the Brownian motion, the back-reaction on the
bath of oscillators needs to be taken into account. In fact,
the heavy quark produces a disturbance that is superimposed to the harmonic motion
of the oscillators, and this disturbance couples to the heavy quark itself causing dissipation.
 
One of the ingredients of this model is the distribution of oscillators, $dN/d\omega$, where $\omega$
denotes the proper frequency of the oscillator. This quantity affects 
the shape of the dissipative kernel: in particular, if $dN/d\omega\propto\omega^2$ then
the kernel is of a Markov type, $\gamma(t-t^\prime)\propto \gamma\delta(t-t^\prime)$ so there is 
no memory; on the other hand, deforming the above distribution it is possible to obtain a kernel
with memory, making possible to study the effect of the latter on physical processes like thermalization.
This model allows to obtain almost all analytical results, albeit 
in the weak coupling regime, permitting to understand 
how the distribution of the bath and memory kernel affect a physical process.

The main difference between our study and those based on generalized Langevin equations
is that in ours
the motion of the oscillators and of the heavy quark is deterministic, while in the Langevin
equations the noise is a random number with a given correlation. Within the present model,
the random nature of the bath comes into the game as soon as we assume a distribution for the
initial conditions of the oscillators; averaging physical quantities over the initial conditions 
amounts to take the ensemble average for the system. The apparent merit of the present 
model is that it is very easy 
to implement a bath that is out-of-equilibrium, which is an important aspect to consider
when studying the diffusion of heavy quarks in the evolving Glasma.
Moreover, the dissipative kernel appears clearly as an effect of the coupling of the heavy quark
to the disturbance it creates in the bath and this might be useful to implement
the dissipative term also in field theories.

The plan of the article is as follows: in Section II we describe the model in detail and derive the evolution
equation for the heavy quark momentum, $P$, as well as define the shape of the dissipative kernel; 
in Section III we solve exactly the evolution equation for $P$ and estimate the thermalization time
and how the memory affects this quantity within the model;
in Section IV we briefly describe  the equilibrium state; finally in Section V we present our summary and conclusions.

 \section{The model}

\subsection{The Hamiltonian}
We consider the classical dynamics of a nonrelativistic particle with mass M interacting with a bath of
$N\gg1$ harmonic oscillators characterized by mass $m$ and frequencies $\omega_k$; we will also introduce a
distribution function for the bath of oscillators, $dN/d\omega$, with normalization
\begin{equation}
N =\int_{-\infty}^\infty\frac{dN}{d\omega}d\omega.\label{eq:normalization_2}
\end{equation} 
In the above equations we have assumed that the oscillator density is an even function of $\omega$. 
As it will be clear soon, $dN/d\omega$ (to be precise, its Fourier transform) 
is the only quantity needed to build up the dissipative kernel,
therefore changing  its analytical form allows to turn from a kernel without a memory to one with memory.
The Hamiltonian is \cite{RUSSO}
\begin{equation}
H=\frac{P^2}{2M} +V(Q)+ \sum_{k=1}^N \left(
\frac{p_k^2}{2m} + \frac{m\omega_k^2 x_k^2}{2} 
\right) -g Q\sum_{k=1}^N x_k + {\cal H}_R,
\label{eq:_renoH}
\end{equation} 
where $M\gg m$ and $Q$ corresponds to the coordinate of the heavy quark.  
The interaction depends on the coupling $g$, which has the dimensions of a cubic energy, that we take to be the same for each oscillator;
in order to mimic a hot QCD medium it would be desirable to have a coupling that depends on the energy scale of the
bath, namely on the temperature: while the implementation of this would be straightforward, we leave it to a future work
because the main calculations on equilibration are unaffected by this further complication.
The term ${\cal H}_R$ in Eq.~\eqref{eq:_renoH} is a renormalization potential~\cite{CHAOS}  that is introduced in the model
in order to balance the effect of the back-reaction on the bath leaving $V(Q)$ the potential: we will specify this term later
and its purpose will appear less obscure as soon as the dynamical equation for $P$ will be derived.
In actual calculations we replace summations over oscillators with an integral: for any $f$ we will use
\begin{equation}
\sum_{k=1}^N f(\omega_k)\rightarrow \int_{-\infty}^{+\infty}\frac{dN}{d\omega}f(\omega) d\omega.\label{eq:replace}
\end{equation}

For a bath in thermal equilibrium the initial conditions for the oscillators 
are distributed randomly according to the standard Boltzmann distribution at temperature $T$, 
\begin{equation}
f(p_{0k},x_{0k}) = e^{-E^p_k/T}e^{-E^x_k/T};\label{eq:therm_dist}
\end{equation} 
where we have put
\begin{eqnarray}
E^p_k &=& \frac{p_{0k}^2}{2m},\\
E^x_k &=& \frac{m\omega^2_k x_{0k}^2}{2},
\end{eqnarray}
which imply $\langle p_{0k}^2\rangle = mT$ and $m\omega_k^2\langle x_{0k}^2\rangle = T$
in agreement with the equipartition theorem.
On the other hand, when we consider an out-of-equilibrium system we introduce a Fermi-Dirac-like distribution, namely
\begin{equation}
\frac{1}{1+e^{(E^p_k - {\cal M})/{\cal T}}}\frac{1}{1+e^{(E^x_k - {\cal M})/{\cal T}}},\label{eq:ooe_dist}
\end{equation}
where ${\cal T}$ and ${\cal M}$ are two parameters with the dimension of energy.  
When ${\cal M}/{\cal T}\gg1$ the out-of-equilibrium distribution becomes a theta-function that has been used
in the literature to mimic the 
gluon distribution in the Glasma \cite{Scardina:2014gxa,Blaizot:2013lga,Blaizot:2011xf}, 
with ${\cal M}$ playing the role of the saturation scale $Q_s$:
for this reason, we call Eq.~\eqref{eq:ooe_dist} as the CGC-like distribution.
The form in Eq.~\eqref{eq:ooe_dist} also keeps $x_{0k}$ and $p_{0k}$ uncorrelated:
while it would be interesting to study the effects of correlation between coordinate and momentum space,
we neglect these for simplicity.
The bath distribution in Eq.~\eqref{eq:ooe_dist} implies 
$\langle p_{0k}^2\rangle = m{\cal T}_\mathrm{eff}$ and $m\omega_k^2\langle x_{0k}^2\rangle = {\cal T}_\mathrm{eff}$
where we have defined the effective temperature of the bath,
\begin{equation}
{\cal T}_\mathrm{eff} = {\cal T}\frac{\mathrm{Li}_{3/2}(-z)}{\mathrm{Li}_{1/2}(-z)},\label{eq:eff_temp_ooe}
\end{equation}
with $z=e^{{\cal M}/{\cal T}}$ and $\mathrm{Li}_s$ denotes the Polylogarithm function of order $s$.
 For example, in the case of a flat distribution ${\cal M}/T\rightarrow \infty$ we find
\begin{equation}
{\cal T}_\mathrm{eff} = \frac{2{\cal M}}{3}.\label{eq:angelica_2}
\end{equation}

Another ingredient of the present model is the distribution of the
frequency of the oscillators, $\omega_k$: we distribute this with 
the density $dN/d\omega$ that will be specified later case by case and that will be normalized according to the
condition in Eq.~\eqref{eq:normalization_2}; we can anticipate that the memory effect of the dissipative kernel
is determined solely by the form of $dN/d\omega$.
Although this should be obvious we remark that we make the assumption of a rigid $dN/d\omega$,
meaning that although we consider the back-reaction of the heavy quark on the bath of oscillators,
our dynamics is such that this back-reaction affects only the motion of the oscillators and not their internal
structure, namely $dN/d\omega$ is unaffected by the presence of the heavy quark.

\subsection{The dynamical equation for P}
The derivation of the dynamical equation for $P$ is straightforward and well known in the literature;
however, in order to emphasize that dissipation comes solely from the back-reaction on the bath of oscillators
we review the main steps of the derivation here; this will also help to clarify the renormalization of the potential
in Eq.~\eqref{eq:_renoH}. 
The starting point is the equation of motion of the oscillator $k$, that can be obtained easily by the Hamilton's equations,
that is
\begin{equation}
\frac{d^2 x_k}{dt} + \omega_k^2 x_k =\frac{g}{ m}Q;\label{eq:ho_1}
\end{equation}
this equation can be easily solved exactly using the retarded Green's function, namely  
\begin{equation}
x_k(t) = x_k^h(t) - \frac{g}{ m} \int_0^t G_R(t-t^\prime) Q(t^\prime) d t^\prime,\label{eq:gf_1}
\end{equation}
where we have defined
\begin{equation}
{\cal G}_R(x) = \theta(x)G_R(x),
\end{equation}
with ${\cal G}_R(x)$ denoting the retarded Green's function of the oscillator,
\begin{equation}
\left(\frac{d^2}{dt^2} + \omega_k^2\right){\cal G}_R(t-t^\prime) = -\delta(t-t^\prime);
\end{equation}
moreover we have defined
\begin{equation}
x_k^h(t) = \alpha_{k}\cos\omega_k t + \beta_k  \sin\omega_k t,
\end{equation}
corresponding to the unperturbed harmonic motion of the oscillator
(namely to the solution of the homogeneous equation),
with $\alpha_k = g x_{0k}$, $m\omega_k\beta_k = gp_{0k}$ and
$x_{0k}$ and $p_{0k}$ denote the initial conditions for coordinate and momentum of each oscillator in the bath.
More generally, $x_k^h(t)$ corresponds to the solution of the homogeneous equation of the oscillator.
In this work we assume  a statistical distribution for these initial conditions, but the dynamics
of the heavy quark as well as that of the oscillators is purely deterministic. 
This approach is very similar to what is done
within the CGC effective theory, in which a random distribution (usually gaussian) 
of color charges is assumed, then the initial color fields are computed
and the evolution is studied by solving the CYM equations which are deterministic; 
within the CGC formalism the ensemble average is taken numerically by averaging the physical quantities over many
different initializations of the color charges. In our work we follow a similar path:
we firstly distribute the initial conditions for the oscillators by virtue of a statistical distribution,
either at equilibrium or out of equilibrium, then the evolution of the system is studied by
the classical equations of motion, and the ensemble average is obtained by averaging the classical solutions
with the statistical distribution of the initial conditions.
We remark that Eq.~\eqref{eq:gf_1} is exact and contains the back-reaction 
induced by the interaction of the heavy quark with the oscillator, the latter being represented by
the convolution of the Green's function with the generalized coordinate $Q$; the analytical form of the Green's function
will be specified later.  

The equation of motion for $P$ is just the second Hamilton's equation that can be read from Eq.~\eqref{eq:_renoH},
namely
\begin{equation}
\frac{dP}{dt}  =-\frac{\partial V}{\partial Q} -\frac{\partial{\cal H}_R}{\partial Q}+  g\sum_k x_k(t).
\end{equation}
Defining the source term, $\xi(t)$, as
\begin{equation}
\xi(t) = g\sum_{k=1}^N x_k^h(t),
\label{eq:rum_disc}
\end{equation}
where $x_k^h(t)$ denotes the harmonic motion of the momentum of the $k^{th}$ oscillator, and taking into account
Eq.~\eqref{eq:gf_1} we can write
\begin{equation}
\frac{dP}{dt} +\frac{ g^2}{ m}\sum_{k=1}^N \int_0^t G_R(t-t^\prime) Q(t^\prime) d t^\prime =
-\frac{\partial V}{\partial Q} -\frac{\partial{\cal H}_R}{\partial Q} + \xi(t).
\end{equation}
The convolution term in the left hand side of the above equation represents the 
coupling of the heavy quark to the perturbation of the harmonic
motions of the oscillators which is induced by the heavy quark itself:
it is therefore a back-reaction term. Note that 
by an integration by parts we can write this convolution as
\begin{equation}
\int_0^t G_R(t-t^\prime) Q(t^\prime) d t^\prime = -F(0) Q(t) + F(t) Q(0) + 
\frac{1}{M}\int_0^t F(t-t^\prime) P(t^\prime) d t^\prime,
\end{equation}
where $F(x)$ is the primitive of $G_R(x)$ and we have taken into account that
$dG_R(x)/dx = -d G_R(t-t^\prime)/dt^\prime$. The full equation for $P$ can thus be written as
\begin{equation}
\frac{dP}{dt} + \frac{g^2}{mM}\sum_{k=1}^N \int_0^t F(t-t^\prime) P(t^\prime) d t^\prime =
-\frac{\partial V}{\partial Q} -\frac{\partial{\cal H}_R}{\partial Q} +
\sum_{k=1}^N F(0) Q(t) - \sum_{k=1}^N F(t) Q(0)+ \xi(t).\label{eq:gto_2}
\end{equation}

The role of ${\cal H}_R$ appears now clearly: 
the term $-F(0) Q(t)$ in Eq.~\eqref{eq:gto_2} corresponds to a harmonic shift of $V(Q)$ induced by
the back-reaction;  taking
\begin{equation}
{\cal H}_R = \frac{1}{2}\sum_k F(0) Q^2\label{eq:ren_HR_5}
\end{equation}
amounts to cancel this shift when the summation over the oscillators is performed,
and $V(Q)$ remains the potential. Using Eq.~\eqref{eq:ren_HR_5} we get
\begin{equation}
\frac{dP}{dt} + \frac{g^2}{mM}\sum_{k=1}^N \int_0^t F(t-t^\prime) P(t^\prime) d t^\prime =
-\frac{\partial V}{\partial Q} - F(t) Q(0)+ \xi(t).\label{eq:gto_3}
\end{equation}
We now make the assumption that $Q(0)=0$: this assumption can be relaxed easily, as it will be clear when we will present
the exact solution of the equation of motion. 
Moreover, here we want to study the motion of the heavy quark in the bath solely therefore we put $V(Q)$ = 0.
In this case we get 
\begin{equation}
\frac{dP}{dt} + \frac{g^2m}{M}\sum_{k=1}^N \int_0^t F(t-t^\prime) P(t^\prime) d t^\prime =
   \xi(t).\label{eq:gto_4}
\end{equation}
This equation looks like a generalized Langevin equation: however, we remark that the term $\xi(t)$ on the right
hand side of the above equation is not a noise but a deterministic source, therefore the motion of the heavy quark
is completely deterministic. An ensemble average will be taken on $P(t)$ averaging over the initial conditions
of the harmonic oscillators; we will prove later that for a Markov process the time correlator of
$\xi(t)$, defined in terms of the ensemble average, is indeed proportional to a $\delta-$function in analogy with a
Gaussian noise used in standard Langevin equations.

The last step is to write explicitly the function $F(x)$: to this end we need the Green's function of the harmonic oscillator,
\begin{equation}
G_R(t-t^\prime) = -\frac{\sin[\omega_k(t-t^\prime)]}{\omega_k};
\end{equation}
the primitive of $G_R$ is easily calculated; defining
the dissipative kernel as 
\begin{equation}
\gamma(t-t^\prime) = \frac{g^2 }{mM} \sum_{k=1}^N \frac{\cos[\omega_k(t-t^\prime)]}{\omega_k^2}.
\label{eq:gamma_disc}
\end{equation}
we can write the dynamical equation for $P$ as
\begin{equation}
\frac{dP}{dt} + \int_0^t \gamma(t-t^\prime) P(t^\prime)dt^\prime = \xi(t).
\label{eq:forP} 
\end{equation}

The derivation of Eq.~\eqref{eq:forP} shows that the effect of the coupling of the heavy quark
to the oscillators appears in two places: on the right hand side of the equation, that describes
the coupling to the unperturbed harmonic motion of the oscillators and that is responsible of momentum diffusion
as we will see shortly,
and in the dissipative term on the left hand side,
that instead describes the coupling of the heavy quark to the perturbation created by its motion in the bath
and is responsible of dissipation:
neglecting this back-reaction amounts to neglect energy loss and the thermalization of the heavy quark
is not achievable.
The back-reaction term  appears as a next-to-leading order interaction, however 
the drag and diffusion coefficients are of the same order in the coupling as we will remind in the next subsection;
nevertheless, we will also provide an estimate of the kinematic regime
in which neglecting the dissipative term is a reasonable approximation. 
In the continuum limit using the replacement~\eqref{eq:replace} we can write Eq.~\eqref{eq:gamma_disc} as
\begin{equation}
\gamma(t-t^\prime) = \frac{g^2 }{2mM} \int_{-\infty}^\infty d \omega
\left(\frac{1}{\omega^2}\frac{dN}{d\omega}\right)
e^{i\omega(t-t\prime)},
\label{eq:gamma_cont_inf}
\end{equation}
which gives the Markov limit $\gamma(t-t^\prime) = 2\gamma\delta(t-t^\prime)$ for $dN/d\omega \propto \omega^2$;
in this case we get
\begin{equation}
\frac{dP}{dt} + \gamma  P = \xi(t). \label{eq:P_markov}
\end{equation}
We notice from Eq.~\eqref{eq:gamma_cont_inf} that $(1/\omega^2)dN/d\omega$ is, besides some multiplicative constant,
nothing but the Fourier transform of the dissipative kernel: it appears clear therefore that the shape of the latter
can be modified by that of the former.
The ensemble average will be taken on the solution of Eq.~\eqref{eq:forP}: practically, this means that we 
solve the deterministic equation of motion~\eqref{eq:forP} for a set of initial conditions of the oscillators,
then we average over these by means of the distributions specified in the previous subsection.

\subsection{The fluctuation-dissipation theorem}
We now show that as long as we use Eqs.~\eqref{eq:therm_dist} and~\eqref{eq:ooe_dist}
to distribute the bath of harmonic oscillators, the dissipative kernel and the correlator $\langle\xi(t)\xi(t^\prime)\rangle$
satisfy the fluctuation-dissipation theorem regardless of the particular $dN/d\omega$ chosen.
To this end, taking into account that from Eqs.~\eqref{eq:therm_dist} and~\eqref{eq:ooe_dist}
it follows that $\langle\alpha_k \alpha_{q}\rangle\propto \delta_{kq}$,
$\langle\beta_k \beta_{q}\rangle\propto \delta_{kq}$, $\langle\alpha_k \beta_{q}\rangle= 0$,
we get
\begin{eqnarray}
\langle\xi(t)\xi(t^\prime)\rangle &=& \frac{1}{2}\sum_k\left\{
\langle \alpha_{k}^2\rangle +  \langle \beta_{k}^2\rangle\right\}\cos[\omega_k(t-t^\prime)]
\nonumber \\
&&+\frac{1}{2}\sum_k\left\{
\langle \alpha_{k}^2\rangle -  \langle \beta_{k}^2\rangle\right\}\cos[\omega_k(t+t^\prime)].
\label{eq:general_xx}
\end{eqnarray}
The above equation is general. If we now make the further assumption that
\begin{equation}
\langle p_{0k}^2\rangle = m^2\omega_k^2\langle x_{0k}^2\rangle,\label{eq:assumption}
\end{equation}
we get
\begin{equation}
\langle\xi(t)\xi(t^\prime)\rangle = \frac{g^2}{m^2}\sum_k \frac{\langle p_{0k}^2\rangle}{\omega_k^2}
\cos[\omega_k(t-t^\prime)],;\label{eq:wedenote222}
\end{equation}
assuming that $\langle p_{0k}^2\rangle$ is independent
on the oscillator, which happens for the distributions used in this work, we can write
\begin{equation}
\langle\xi(t)\xi(t^\prime)\rangle = \frac{g^2\langle p_{0}^2\rangle}{m^2}\sum_{k=1}^N 
\frac{\cos[\omega_k(t-t^\prime)]}{\omega_k^2}
,\label{eq:wedenote}
\end{equation}
where now $\langle p_{0}^2\rangle$ corresponds to the common average initial squared momentum
of the oscillators. 

A comparison of Eq.~\eqref{eq:wedenote} with Eq.~\eqref{eq:gamma_disc} shows that within this model
the fluctuation-dissipation theorem is satisfied in the form
\begin{equation}
\gamma(t-t^\prime) = \frac{m}{M\langle p_{0}^2\rangle}\langle\xi(t)\xi(t^\prime)\rangle.\label{eq:FD_general}
\end{equation}
This relation is valid regardless of the particular form of $dN/d\omega$, as long as the condition~\eqref{eq:assumption}
is satisfied.

We remark that in order to satisfy the fluctuation-dissipation theorem the assumption~\eqref{eq:assumption}
is crucial: if this condition is violated then the second term on the right hand side of Eq.~\eqref{eq:general_xx} 
does not necessarily cancels and this might lead to the violation of the theorem. 
This might happen for example if in the initial condition the degrees of freedom of the oscillators are thermalized
at a different temperature: this situation represents a different out-of-equilibrium condition, and
in presence of interactions among the modes of the oscillators this condition should be removed 
by the dynamical evolution.
While we leave this possibility
to a future investigation, in this study the distributions of the oscillators always satisfy the condition~\eqref{eq:assumption}
therefore the aforementioned term cancels and the fluctuation-dissipation theorem is always satisfied. 

\subsection{The pure diffusive motion\label{sec:pdm}}

We now show that in this model the heavy quark experiences a pure diffusion
over the initial value of momentum in case the dissipative term is neglected, in case of a Markov process; indeed, 
in this case the equation of motion is solved by
\begin{equation}
P(t) = P(0) + \int_0^t dt^\prime\xi(t^\prime);
\label{eq:opop}
\end{equation}  
squaring and taking the ensemble average leads to
\begin{eqnarray}
\langle P(t)\rangle &=& P(0),\\
\left\langle P(t)^2 \right\rangle &=&  P(0)^2  + \frac{2 g^2 {\cal E}}{m}
\int d\omega\frac{dN}{d\omega}\frac{1-\cos\omega t}{\omega^4},
\end{eqnarray}
where we have put
${\cal E}= T$ for the system in thermal equilibrium and ${\cal E} = {\cal T}_\mathrm{eff}$ for CGC-like distribution.
The time dependence of momentum spreading comes from the integral over the frequencies of the oscillators on the
right hand side of the above equation; for the case of a Markov process this is easy to compute because this corresponds
to $dN/d\omega\propto\omega^2$ which implies
\begin{equation}
\left\langle \frac{P(t)^2}{2M} \right\rangle -  \frac{P(0)^2}{2M}   = 2Dt.\label{eq:elia_cra}
\end{equation}
where we have defined the diffusion coefficient for the kinetic energy as
\begin{equation}
D = \frac{ \pi g^2 {\cal E}}{2mM}\left(\frac{1}{\omega^2}\frac{dN}{d\omega}\right).
\label{eq:D_C_E}
\end{equation}
The above equations describe a standard diffusion with momentum spreading linearly
with time. 

We can compare the diffusion coefficient
with the drag in the Markov limit: indeed,
from Eq.~\eqref{eq:gamma_cont_inf} we get
\begin{equation}
\gamma = \frac{\pi g^2}{2mM}\left(\frac{1}{\omega^2}\frac{dN}{d\omega}\right),\label{eq:Dr_C_E}
\end{equation}
and comparing this with Eq.~\eqref{eq:D_C_E} we find
\begin{equation}
\frac{D}{\gamma} =  {\cal E};
\label{eq:FDT_ooe}
\end{equation}
this corresponds to the fluctuation-dissipation theorem specialized to the case of a Markov process.

\subsection{The gaussian dissipative kernel}
In this study we consider a dissipative kernel with a memory: 
as already specified above, a Markov kernel (namely, a kernel without memory)
corresponds to the limit $\gamma(t-t^\prime) = 2\gamma\delta(t-t^\prime)$
and according to Eq.~\eqref{eq:gamma_cont_inf} this can be implemented by means of 
$dN/d\omega \propto \omega^2$. In order to keep expressions that are easy to manipulate
we consider a simple gaussian form 
\begin{equation}
\frac{dN}{d\omega} = \frac{2N}{a^3\sqrt{\pi}}\omega^2 e^{-\omega^2/a^2},\label{eq:gauss_dN}
\end{equation}
which is normalized according to Eq.~\eqref{eq:normalization_2}; in this equation
we introduce the parameter $a$, which carries the dimension of energy,
that regulates the shape of the kernel in the frequency space.
Using Eq.~\eqref{eq:gauss_dN} we get from Eq.~\eqref{eq:gamma_cont_inf}
\begin{equation}
\gamma(t-t^\prime) = \frac{2\alpha^2}{M}\Phi_a(t-t^\prime),
\label{eq:gamma_cont_333_NEWnew}
\end{equation}
where
\begin{equation}
\Phi_a(t-t^\prime) = \frac{a}{2\sqrt{\pi}}\exp\left(-\frac{a^2(t-t^\prime)^2}{4}\right),
\label{eq:new_fa}
\end{equation}
and
\begin{equation}
\alpha^2 = \frac{g^2 N }{a^3 m}\sqrt{\pi}.\label{eq:eff_coupl_ggg}
\end{equation}
This form has the advantage to identify easily the part of the kernel that gives the $\delta-$function in the
$a\rightarrow\infty$ limit, since $\Phi_a(x)\rightarrow\delta(x)$ in this limit. The form given in Eq.~\eqref{eq:gamma_cont_333_NEWnew}
also suggests that the proper way to take the $a\rightarrow\infty$ limit is to keep $g^2 N/a^3$ fixed, so that the overall
constant in Eq.~\eqref{eq:gamma_cont_333_NEWnew} is unchanged as we continuously deform the dissipative kernel from one with memory
to the Markov one $\gamma(s)=2\gamma\delta(s)$. In addition to this, we notice that Eq.~\eqref{eq:gamma_cont_333_NEWnew}
suggests we can define an effective coupling,
that contains the information about the bath of oscillators and is independent on the mass of the heavy quark;
as a consequence, once we fix $M$ we can study the effect of a continuous deformation of the dissipative kernel
by keeping $\alpha^2$ fixed: this limiting procedure gives meaningful results both if we take the $N\rightarrow\infty$
and the $a\rightarrow\infty$ limits.

\section{The exact solution and equilibration time}
The dynamical equation for $P$, Eq.~\eqref{eq:forP}, is an integro-differential equation that can be solved
analytically by means of Laplace transform. We present this solution here, then we discuss how we can use this
solution in order to define a thermalization time and how we can estimate this for a given dissipative kernel.

\subsection{Exact solution and equilibration time}

In order to solve Eq.~\eqref{eq:forP} we introduce the Laplace transform, namely 
\begin{equation}
{\cal L}_s(f)=F(s)=\int_0^\infty e^{-st}f(t)dt;
\end{equation}
it is then straightforward to write Eq.~\eqref{eq:forP} as  
\begin{equation}
s{\tilde P}(s)-P(0)+\Gamma(s){\tilde P}(s)=\Xi(s);\label{eq:from_above}
\end{equation}
here $\Xi(s)$ and $\Gamma(s)$ correspond to the Laplace transform of the source term and of the dissipative kernel
respectively.
From Eq.~\eqref{eq:from_above} we can write the solution in the time domain as the inverse Laplace transform 
of $\tilde{P}(s)$, that is
\begin{equation}
P(t)=\frac{1}{2\pi i}\int_{\sigma-i\infty}^{\sigma+i\infty}\frac{\Xi(s)+P(0)}{s+\Gamma(s)}e^{st}ds,
\label{eq:exact_solu_t}
\end{equation}
where $\sigma$ is a real constant that exceeds the real part of all the singularities of the integrand.
Equation~\eqref{eq:exact_solu_t} corresponds to the exact solution of Eq.~\eqref{eq:forP}: in principle
as soon as the dissipative kernel and the source are given, one can obtain the evolution of $P$
by performing the integral above; then, an ensemble average can be taken by averaging over the
initializations of the oscillators in the bath.

Before specializing the solution~\eqref{eq:exact_solu_t} to some specific dissipative kernel and source,
we notice that we can split this solution into two parts:
\begin{equation}
P(t)=\frac{P(0)}{2\pi i}\int_{\sigma-i\infty}^{\sigma+i\infty}\frac{1}{s+\Gamma(s)}e^{st}ds
+
\frac{1}{2\pi i}\int_{\sigma-i\infty}^{\sigma+i\infty}\frac{\Xi(s)}{s+\Gamma(s)}e^{st}ds
\label{eq:exact_solu_t_2};
\end{equation}
in this equation, $P(0)$ is the initial value of the heavy quark's momentum,
therefore the first addendum on the right hand side is the only piece of solution that contains information on this
initialization. Depending on the distribution of the poles of the integrand, the magnitude of this term 
can grow up or be damped exponentially, as well as oscillate in time. If all the poles 
have negative and nonzero real part then this first addendum will decay exponentially,
implying that after some time the heavy quark loses memory of its initial condition
and its properties at asymptotic time will depend only on those of the bath in which the quark itself propagates. 
This offers a criterion for thermalization, or more generally for equilibration, of the heavy quark
with the medium: indeed, we can define the thermalization time as the time needed for the quark to 
forget about its initial condition. This time, that we denote by $\tau_\mathrm{eq}$, 
can be computed by studying the distribution of the zeros
of $s+\Gamma(s)$ in the complex plane; 
assuming poles $s=-\kappa_r + i\kappa_i$, with $\kappa_r$ and $\kappa_i$ corresponding to the real and imaginary part
of $s$ respectively, and assuming that all the $\kappa_r$ are positive, we can identify
$\tau_\mathrm{eq}$ with the smallest of the $\kappa_r$.
We notice that using this definition, $\tau_\mathrm{eq}$ depends only on the shape of the dissipative kernel 
in frequency space and not on the initial conditions of the oscillators bath:
given a density $dN/d\omega$, the equilibration time will depend only on this as well as on the coupling and the masses
of the heavy quark and of the oscillators, but not on the distribution of the oscillators in the bath.

\subsection{Equilibration time for the gaussian kernel}
We now specialize the solution~\eqref{eq:exact_solu_t} to the case of the gaussian dissipative kernel 
Eq.~\eqref{eq:gamma_cont_333_NEWnew}.
After writing down the exact solution, we will use it to determine the equilibration time as defined
in the previous subsection as a function of the shape parameter of the kernel.
The Laplace transform of the kernel is
\begin{equation}
\Gamma(s) = \frac{\alpha^2}{M}
e^{\frac{s^2}{a^2}}E_a\left(s/a\right),\label{eq:Gamma_s}
\end{equation}
where $N$ is the number of oscillators in the bath and we have put
\begin{equation}
E_a(x)=\frac{2}{\sqrt{\pi}}\int_x^\infty e^{-t^2}dt;
\end{equation}
we remind that we obtain the Markov limit when we take $a\rightarrow\infty$ in the above equation.
The Laplace transform of Eq.~\eqref{eq:rum_disc} is given by 
\begin{equation}
\Xi(s)=\sum_{k=1}^N\left(\frac{\alpha_ks+\beta_k\omega_k}{s^2+\omega_k^2}\right).
\end{equation}
Therefore, we can write the exact solution in the form
\begin{equation}
P(t)=\frac{P(0)}{2\pi i}\int_{\sigma-i\infty}^{\sigma+i\infty}\frac{1}{s+\Gamma(s)}e^{st}ds+
\frac{1}{2\pi i}\sum_{k=1}^N\int_{\sigma-i\infty}^{\sigma+i\infty}
\left(\frac{\alpha_k s+\beta_k\omega_k}{s^2+\omega_k^2}\right)\frac{1}{s+\Gamma(s)}e^{st}ds.
\label{eq:da_ref_3456}
\end{equation}
A quick calculation of the residues at infinity of the
integrands in Eq.~\eqref{eq:da_ref_3456} at $t=0$,
which correspond to the integrals at $t=0$, shows that the solution~\eqref{eq:da_ref_3456}
is consistent with $P(t=0) = P(0)$: the second addendum in the right hand side of the equation 
vanishes at $t=0$ while the first one is simply equal to $P(0)$.

As discussed in the previous subsection, in order to estimate the equilibration time we have to study the
distribution of the poles of the integrands in the right hand side of Eq.~\eqref{eq:exact_solu_t_2};
the equation $s+\Gamma(s)=0$ with $\Gamma(s)$ in Eq.~\eqref{eq:Gamma_s}
admits both negative real, $s^\ell=-\chi^\ell$, 
and a set of infinite conjugate complex solutions $s^n=-\kappa^n_r\pm i\kappa^n_i$ with $\kappa^n_r>0$. 
Besides, there are conjugate poles $s_k=\pm i\omega_k$ in the second integral. 
We have been able to find the explicit solutions to the equation $s+\Gamma(s)=0$ analytically only for small $\alpha^2$
in the large  $a$ limit, therefore for a bath that is not very far from the Markov limit:
these can be found by using the Newton-Raphson method iterated up to the second order
(we checked that increasing the order of the iteration does not change the result in the weak coupling
limit). These solutions are given by
\begin{eqnarray}
\chi^1 &=& \frac{\alpha^2 }{M} +\frac{2}{a\sqrt{\pi}}\frac{\alpha^4}{M^2} 
+ O\left(\frac{\alpha^6}{M^3}\right), \\
\kappa_r^n &=& na +  \frac{\alpha^2 }{M} +\frac{2}{a\sqrt{\pi}}\frac{\alpha^4}{M^2}
+ O\left(\frac{\alpha^6}{M^3}\right),~~~n=1,2,\dots
\end{eqnarray}
with $\kappa_i^n = -na$. Clearly the smallest zero is $\chi^1$ and in the limit of large $a$
the $\kappa_r^n$ is very large, which means that the contribution of these poles is suppressed 
as soon as $1/\kappa_r^1 < t $ and in this time range we can limit
ourselves to consider only $\chi^1$.
In this limit, by a straightforward application of the residue theorem we get
\begin{eqnarray}
P(t)&=&P(0)  e^{-\chi^1 t} 
\nonumber\\
&&+  e^{-\chi^1 t} \sum_{k=1}^N
\frac{-\alpha_k\chi^1+\beta_k\omega_k}{(\chi^1)^2+\omega_k^2} 
\nonumber \\
&&+\sum_{k=1}^N
\left[\frac{i\alpha_k\omega_k+\beta_k\omega_k}{2i\omega_k}\frac{e^{i\omega_kt}}{i\omega_k+\Gamma(i\omega_k)} 
 + \mathrm{c.c.}\right],
\label{eq:full_gaussian}
\end{eqnarray}
where c.c. denotes complex conjugation; the last line in the above equation corresponds to the contribution
of the poles $s=\pm i\omega_k$ and represents the 
heavy quark momentum at very large time while the first two terms are important only in the transient region.

As anticipated, a part of the solution~\eqref{eq:full_gaussian} 
depends on the initial condition, see the first line in the right hand side
of the above equation.  
We identify $\chi^1$ with the inverse of the equilibration time, getting
\begin{equation}
\tau_\mathrm{eq} = \frac{M}{\alpha^2}
\left(1 - \frac{2\alpha^2}{a \sqrt{\pi} M } 
+ O(\alpha^4/a^2 )\right).\label{eq:eq_ti_1}
\end{equation}
The above equation shows that the within this model,
the effect of the 
memory in the dissipative kernel is to lower the equilibration time of the heavy quark
when the effective coupling $\alpha^2$ is kept fixed.
We have been unable to find a detailed explanation of why this happens:
our naive interpretation is that adding memory to the dissipative kernel amounts to lower the value of $a$ and,
because of Eqs.~\eqref{eq:gauss_dN} and~\eqref{eq:eff_coupl_ggg}, keeping $\alpha^2$ fixed 
implies that the density of oscillators at a given frequency $\omega$ is larger,
resulting in a more efficient interaction with the medium and in a faster thermalization.
To this end we remark that we have obtained Eq.~\eqref{eq:eq_ti_1} using a gaussian kernel: it will be interesting in the future
to investigate further analytical forms of the dissipative kernel in order to verify if the
lowering of the equilibration time is more general or if it is related to the specific form of the
kernel used in our work.  
Furthermore, it will be interesting to study the impact of memory effect 
on heavy quark observables at Relativistic Heavy Ion Collider (RHIC) and  Large
Hadron Collider (LHC) energies, as well as on heavy quark thermalization time, which 
is neglected in several recent calculations~\cite{vanHees:2005wb, vanHees:2007me, Scardina:2017ipo, Das:2015ana, 
Alberico:2013bza, Cao:2013ita, He:2013zua}. It can be mentioned that memory effect have an important influence on the 
final dilepton yields~\cite{Schenke:2006uh} and radiative energy loss~\cite{Michler:2009dy}. This indicates  
an analysis  which describes the data correctly demands the proper inclusion of memory 
effects (for more see also~\cite{Greiner:1994xm, Murase:2013tma, Kapusta:2014dja,Schmidt:2014zpa, Wesp:2014xpa}).

We notice that the explicit distribution of the bath
does not affect the equilibration time~\eqref{eq:eq_ti_1}: indeed, it is only the shape of the dissipative kernel
that affects the thermalization.
We also notice that although the increase of the average energy per oscillator is fairly negligible, since
the energy lost by the heavy quark during the thermalization is distributed to $N$ oscillators
with $N$ large, the thermalization produces entropy because the information stored in the initial
condition becomes irrelevant after the thermalization time \cite{information}. 

We can combine the results found in this section with those of Section~\ref{sec:pdm}
to give a rough estimate of the kinematic regime in which the motion of the heavy quark
is dominated by diffusion. Indeed, the drag would correspond to a shift of the initial heavy quark momentum
by an amount
\begin{equation}
|P(t)^2 -P(0)^2|_\mathrm{drag} \approx 2P(0)^2 \gamma t,\label{eq:mia_1}
\end{equation}
where we have put $\gamma = 1/\tau_\mathrm{eq}$. On the other hand,  diffusion amount to a dispersion
of the kinetic energy around its initial value, see Eq.~\eqref{eq:elia_cra}; using the fluctuation-dissipation theorem 
in the form of Eq.~\eqref{eq:FDT_ooe}, which is valid both for the bath in thermal equilibrium and for the
CGC-like bath, we can write
\begin{equation}
|P(t)^2 -P(0)^2|_\mathrm{diffusion}  = 4M\gamma{\cal E}t.\label{eq:mia_2}
\end{equation}
Taking the ratio of the last two equations we find
\begin{equation}
\frac{|P(t)^2 -P(0)^2|_\mathrm{drag}}{|P(t)^2 -P(0)^2|_\mathrm{diffusion}} 
\approx \frac{1}{{\cal E}}\frac{P(0)^2}{2M}.\label{eq:anna_bionda}
\end{equation}
Loosely speaking, the above equation shows that the drag can be neglected with respect 
to the diffusion as long as the initial kinetic energy of the heavy quark is small in comparison
with the average energy of the bath. For a bath in thermal equilibrium this average energy is just the temperature;
for a CGC-like bath we have from Eq.~\eqref{eq:angelica_2}
$2{\cal E} \approx {\cal M} \approx Q_s$ with $Q_s$ representing the saturation scale.

A purely diffusive process has been observed in~\cite{Ruggieri:2018rzi} where the propagation of heavy quarks in 
the evolving Glasma
has been studied neglecting the effect of this propagation on the background gluon field;
this is called the probe approximation since the color current carried by the 
heavy quark is supposed to not affect the gluon field.
From the model discussed in this work we can interpret the lack of back-reaction as the lack of a drag coefficient
in the equations of motion of the heavy quarks. 
While we recognize the need of a full numerical solution of the problem of the propagation of heavy quarks in the evolving
Glasma
we can use the present model, and in particular
the estimate~\eqref{eq:anna_bionda}, 
to state that the pure diffusive approximation should work well
 as long as the initial kinetic energy of the heavy quark is small in comparison with 
 the saturation scale. For example, 
taking $Q_s \approx 2$ GeV and
an initial momentum $P(0) = 2$ GeV, which is a fair estimate of the initial average transverse momentum
of heavy quarks produced via perturbative QCD \cite{Ruggieri:2018rzi},
from Eq.~\eqref{eq:anna_bionda} we read the ratio between drag and diffusion
being approximately $0.67$ for the $c$ quark and $\approx 0.22$ for the $b$ quark,
meaning that the pure diffusive approximation in this example works pretty well for the $b$ quark
but is marginal for the $c$ quark; 
for higher momenta the back-reaction need to be considered
in order to have a correct description of the evolution of the heavy quark in the dense gluon system.

\section{The equilibrium state}
In this section we analyze the solution of the equation of motion of $P$ at very large time.
This is important because it gives information on the equilibrium state of the heavy quark in the bath of oscillators.
Moreover, it is useful to understand how the non-Markov dissipative kernel, as well as the
out-of-equilibrium oscillators bath, affect the large time state of the heavy quark. 
From Eqs.~\eqref{eq:full_gaussian} and~\eqref{eq:da_ref_3456} we notice that only the poles $s=\pm i\omega_k$
give a contribution at large time since the other poles lead to an exponential decay of the solution;
therefore, for very large time we can write
\begin{equation}
P(t)=
\frac{1}{2\pi i}\sum_{k=1}^N\int_{\sigma-i\infty}^{\sigma+i\infty}
\left(\frac{\alpha_k s+\beta_k\omega_k}{s^2+\omega_k^2}\right)\frac{1}{s+\Gamma(s)}e^{st}ds,
\label{eq:da_ref_3456_lt}
\end{equation}
and taking into account only the relevant poles we get
\begin{eqnarray}
P(t)&=&
\sum_{k=1}^N
\left(\frac{i\alpha_k +\beta_k }{2i}\frac{e^{i\omega_kt}}{i\omega_k+\Gamma(i\omega_k)} 
 + \mathrm{c.c.}\right).
\label{eq:full_gaussian_infinite}
\end{eqnarray}
We note that although the poles corresponding to $s+\Gamma(s)=0$ do not contribute, the dissipative kernel
still appears explicitly in Eq.~\eqref{eq:full_gaussian_infinite}. 

We use  Eq.~\eqref{eq:full_gaussian_infinite} to compute the average kinetic energy of the heavy quark.
In order to do this we need to square the right hand side of the equation, then take the ensemble average
with the distributions~\eqref{eq:therm_dist} or~\eqref{eq:ooe_dist}, then sum over the oscillators. This is a pretty straightforward
procedure which is not enlightening, therefore we skip the few mathematical steps and write down the final result:
\begin{eqnarray}
\langle P^2\rangle_\infty &=& \frac{1}{2}\sum_k \frac{\langle\alpha_k^2\rangle + \langle\beta_k^2\rangle}
{(\Gamma^\star - i\omega_k)(\Gamma + i\omega_k)} \nonumber\\
&& + \frac{1}{4}\sum_k {\cal F}_k(t)
\frac{(\langle\alpha_k^2\rangle - \langle\beta_k^2\rangle)
}
{(\Gamma^\star - i\omega_k)^2(\Gamma + i\omega_k)^2},
\label{eq:eva_esa}
\end{eqnarray}
where we have put $\Gamma=\Gamma(i\omega_k)$, $\Gamma^* = \Gamma(-i\omega_k)$ and 
\begin{equation}
{\cal F}_k(t) = e^{-2i\omega_k t}(\Gamma + i\omega_k)^2 + e^{+2i\omega_k t}(\Gamma^\star - i\omega_k)^2.
\end{equation}
We briefly notice that the second term in the right hand side of  Eq.~\eqref{eq:eva_esa}
would contribute only if $\langle\alpha_k^2\rangle \neq \langle\beta_k^2\rangle$;
the evaluation of this term turns out to be very simple in the case of a Markov kernel in which $\Gamma(s)$ is a constant. 
We will not comment further on this term in the present work, although in principle it would be interesting to study 
a situation in which $\langle\alpha_k^2\rangle \neq \langle\beta_k^2\rangle$ since this would correspond to a different
kind of out-of-equilibrium bath, in which the different degrees of freedom thermalize at a different temperature.

For the kind of baths considered in this work we have $\langle\alpha_k^2\rangle = \langle\beta_k^2\rangle = g^2{\cal E}/(m\omega_k^2)$
with ${\cal E}= T$ for the system in thermal equilibrium and ${\cal E} = {\cal T}_\mathrm{eff}$ for CGC-like distribution,
therefore only the first term on the right hand side of Eq.~\eqref{eq:eva_esa} gives a contribution:
\begin{equation}
\langle P^2\rangle_\infty  =  \frac{g^2{\cal E}}{ m} \int_{-\infty}^{+\infty} \frac{1}{\omega^2}\frac{dN}{d\omega}
\frac{d\omega}
{(\Gamma^\star - i\omega)(\Gamma + i\omega)}.
\end{equation}
We have been able to perform the above integral only in the limit of large $a$, in which we take the asymptotic expansion
of the integrand up to the order $1/a$: 
\begin{equation}
\left\langle \frac{P^2}{2M}\right\rangle_\infty = \frac{{\cal E}}{2}
\left[
1+\frac{\alpha^2}{aM\sqrt{\pi}}
\right].\label{eq:bourbon}
\end{equation}
The above equation shows that for a kernel without a memory, the heavy quark equilibrates to the 
average kinetic energy of the bath with a temperature that is equal to the bath's temperature if the bath
is in thermal equilibrium, or equal to the effective temperature if the bath has a CGC-like distribution.
The presence of the memory, which corresponds to the term $\propto 1/a$ in Eq.~\eqref{eq:bourbon},
leads to a higher value of the average kinetic energy.  Clearly, this does not correspond to a measurable increase of the average
energy of the bath: the heavy quark has lost part of its initial energy, $\Delta E$, during the thermalization process, 
and this $\Delta E$ has been distributed to the bath so the average energy of each oscillator has increased of the
amount $\Delta E/N$ which is negligible in the $N\rightarrow\infty$ limit. Nevertheless, the entropy has increased
during thermalization because of the loss of information on the initial condition, as already specified above. 
Since in presence of the memory the kinetic energy of the heavy quark is larger than the one found without a memory,
we conclude that the main effect of the memory is that of decreasing the energy loss.

\section{Summary and Conclusions}
We have considered a simple model for the diffusion of heavy quarks in a hot bath, the latter being modeled by an ensemble
of harmonic oscillators distributed accorded to either a thermal distribution or to an
out-of-equilibrium CGC-like distribution with
a saturation scale. The use of the latter is motivated by the recent increased interest of studying the propagation
of heavy quarks in a dense medium of out-of-equilibrium gluons that can be produced in high energy nuclear collisions;
therefore, this model might help to shed light on the involved dynamics of the early stages of these collisions avoiding the complications
of solving the non-abelian equations of motion of the gluon background coupled to those of the heavy quarks.
Of course, we do not pretend that this model can replace the study of the full problem: we are well aware that
a full implementation of the gluon + quarks dynamics is very necessary to make quantitative predictions,
indeed this is what we have started in \cite{Ruggieri:2018rzi};
nevertheless, the simple model studied here can help to clarify at least the qualitative picture,
suggesting what gives what in the interplay between the heavy quarks and the bath, hence
suggesting how to improve the present calculations in order to get a more complete picture. 
This model of diffusion is well known in the literature since it has been studied in the context of 
dissipation in quantum mechanics~\cite{RUSSO,QHO,QHO2,CHAOS}.  

Within this model it is possible to study the process of thermalization of heavy quarks,
in particular how this depends on the specific form of the distribution used for the bath as well as on the coupling and the
heavy quark mass.  Moreover, it is possible to include memory effects in the dissipative kernel, $\gamma(t-t^\prime)$, and study how this
affects the thermalization. The memory can be implemented easily by choosing properly the distribution of oscillators, $dN/d\omega$,
which appears in the very definition of $\gamma(t-t^\prime)$: for the model considered here,
a Markov kernel $\gamma(t-t^\prime) = 2\gamma\delta(t-t^\prime)$
can be obtained by assuming $dN/d\omega\propto\omega^2$. We have studied memory effects
by introducing continuous deformations of the Markov kernel via a gaussian distribution, which helps to keep mathematical
expressions simple at the same time allows for a smooth transition from the kernel without to a kernel with memory.
Therefore, the two main merits of this model are the ability to implement both memory effects and an out-of-equilibrium
dense bath of oscillators.

We have remarked that in order to obtain the dissipative term in the equation of motion of the heavy quark, namely the drift, 
it is necessary to include the back-reaction on the bath of oscillators. Indeed, from the dynamical equation it is clear that
besides the coupling of the quark to the unperturbed oscillators, which is responsible of diffusion,
there is also a coupling of the heavy quark to the perturbation induced by the latter on the bath, and this is responsible
of the dissipation. Moreover, we have proved that within the assumption 
$\langle p_{0k}^2\rangle = m^2\omega_k^2\langle x_{0k}^2\rangle$ the fluctuation-dissipation theorem
is satisfied both for the bath at thermal equilibrium and for the CGC-like one,
regardless of the shape of the distribution of oscillators.

The solution of the equation of motion for the heavy quark momentum, $P$,
can be obtained easily. From this we have defined the equilibration time
as the time necessary for the system to lose information about its initial condition.
It has been possible to compute analytically the thermalization time by studying the distribution of the zeros
of $s+\Gamma(s)$ in the complex plane, where $\Gamma(s)$ denotes the Laplace transforme of the
dissipative kernel: we select the zero with the smallest negative real part, $-\chi$, and define
the inverse of the thermalization time as $1/\tau_\mathrm{eq}=\chi$.
We have found that $dN/d\omega$ is the only part of information about the bath that
affects the thermalization time: the distribution of the degrees of freedom,
$\langle p_{0k}^2\rangle$  and $\langle x_{0k}^2\rangle$, does not affect the thermalization time.
We have also found that keeping the strength of the effective coupling fixed, and deforming 
$dN/d\omega$ in order do have a kernel with memory, the thermalization time decreases.
We have understood this as a consequence of the larger density of oscillators, which in turn
results in a more efficient way to dissipate energy. We remark that we have obtained this result
using a specific analytical form of $dN/d\omega$, namely a gaussian, therefore in this moment we are unable
to make a general statement on the effect of the memory on the thermalization time.
In the future it will be interesting to study more kernels with memory to check whether 
our finding is specific to the gaussian form used or is more general,
even out of the context of heavy quarks dynamics.

The work presented here has been partly inspired by \cite{Ruggieri:2018rzi} in which we have studied the diffusion of heavy quarks
in an evolving Glasma background. In \cite{Ruggieri:2018rzi} we have used the probe approximation,
which amounts to neglect the back-reaction of the colored current carried by the heavy quarks on the gluon background,
and a diffusion-without-drag has been found to be the main characteristics of the propagation of the heavy quarks
in this medium.
We are able to understand why in \cite{Ruggieri:2018rzi} a lack of drift has been observed:
indeed, 
from the equation of motion for the momentum of the heavy quark we notice that
dissipation arises from the coupling of the heavy quark to the perturbation induced by the quark itself on the
bath of oscillators.  Therefore, in order to observe dissipation, the implementation of the back-reaction
on the gluon background seems necessary. Nevertheless, we can use this model to estimate roughly the kinematic regime
in which the diffusion-without-drag is a fair approximation. Indeed, from the fluctuation-dissipation theorem that we
have derived for the CGC-like distribution we have found that the relative shift of the kinetic energy due to drag and diffusion
is just given by $\approx K_0/{\cal E}$, where $K_0$ denotes the initial
(non-relativistic)  kinetic energy of the heavy quark and 
${\cal E}$ is the energy-per-oscillator of the bath: for a system in thermal equilibrium ${\cal E} = T$ while
for the CGC-like distribution we can take ${\cal E}  \approx Q_s$ where $Q_s$ is the saturation scale.
Therefore our model suggests that as long as the initial energy of the heavy quark is smaller than the
average energy of the bath, the pure diffusive motion is not a bad approximation.

There are numerous possible studies that can be done in the future. As mentioned above, it will be interesting
to explore different shapes of the dissipative kernel.
It seems also interesting to investigate the medium-induced interaction between heavy quarks,
that might affect the thermalization time as well as observables like particle-particle correlations.
Besides, it is of a certain interest to consider more general baths in which
$\langle p_{0k}^2\rangle \neq m^2\omega_k^2\langle x_{0k}^2\rangle$:
this would imply that the different degrees of freedom of the bath equilibrate at a different temperature,
therefore an interesting point to study would be to include self-interactions in the bath in order to study
its equilibration, besides the propagation of the heavy quark in this dynamical bath.
In addition to these problems, another aspect that is worth of an investigation is that of 
a correlation between the different degrees of freedom of the bath.  
We intend to explore the impact of memory on heavy quark dynsmics and its thermalization time at RHIC and 
LHC energies, which people usually ignored in earlier calculations, in a forthcoming article as a application
of this present work.

\section*{ACKNOWLEDGEMENTS}
The authors acknowledge  Gabriele Coci,  Vincenzo Greco, Lucia Oliva, John Petrucci, Salvatore Plumari and in particular 
Xingbo Zhao for inspiration,
useful discussions and comments on the first version of this article.
The work of M. R. and S. K. D.  is supported by the National Science Foundation of China (Grants No.11805087 and No. 11875153)
and by the Fundamental Research Funds for the Central Universities (grant number 862946).

\end{document}